\documentclass[a4paper]{article}

\usepackage{icrc2013}

\title{Gamma-ray constraints on decaying Dark Matter}

\shorttitle{Gamma-ray constraints on decaying Dark Matter}

\authors{
E. Moulin$^{1}$,
M. Cirelli$^{2}$,
P. Panci$^{3}$,
P. Serpico$^{4}$,
A. Viana$^{1,5}$
}

\afiliations{
$^1$ Commissariat \`a l'Energie Atomique et aux Energies Alternatives, Institut de recherche sur les lois fondamentales de l'Univers, 
Service de Physique des Particules, Centre de Saclay, F-91191 Gif-sur-Yvette, France \\
$^2$ Institut de Physique Th\'eorique, Centre National de la Recherche Scientifique, 
Unit\'e de Recherche Associ\'ee 2306 \& Commissariat \`a l'Energie Atomique et aux Energies Alternatives/Saclay, F-91191 Gif-sur-Yvette, France \\
$^3$ CP3-Origins and the Danish Institute for Advanced Study DIAS, University of Southern Denmark, 
Campusvej 55, DK-5230 Odense M, Denmark \\
$^4$ Laboratoire dÕAnnecy-le-Vieux de Physique Th\'eorique, Universite« de Savoie, Centre National de la Recherche ScientiÞque, 
B.P.110, Annecy-le-Vieux F-74941, France  \\
\scriptsize{
$^{5}$ now at: Max-Planck-Institut f \"ur Kernphysik, P.O. Box 103980, D 69029 Heidelberg, Germany
}
}
\def\FERMI{{\sf Fermi}}
\def\PAMELA{{\sf PAMELA}}
\def\HESS{{\sf H.E.S.S.}}
\def\MAGIC{{\sf MAGIC}}
\def\CTA{{\sf CTA}}
\def\ICECUBE{{\sf Icecube}}

\email{emmanuel.moulin@cea.fr}

\abstract{New bounds on decaying Dark Matter (DM) are derived from the $\gamma$-ray measurements of (i) the isotropic residual (extragalactic) background by \FERMI\ and (ii) the Fornax galaxy cluster by \HESS. We find that those from (i) are among the most stringent constraints currently available, for a large range of DM masses and a variety of decay modes, excluding half-lives up to $\sim$10$^{26}$ to few 10$^{27}$ seconds. In particular, they rule out the interpretation in terms of decaying DM of the e$^{\pm}$ spectral features in \PAMELA, \FERMI\ and \HESS, unless very conservative choices are adopted. We also discuss future prospects for \CTA\ bounds from Fornax which, contrary to the present \HESS constraints of (ii), may allow for an interesting improvement and may become better than those from the current or future extragalactic \FERMI\ data.}


\begin{document}
\maketitle

\section{Introduction}
The possibility that Dark Matter (DM), which constitutes most of the matter in the Universe, consists of a particle that actually decays on a very long time scale has attracted much attention lately. This is because the decay time scale $\tau_{\rm dec}$ can be taken to be `short' enough that the decay products give signals in current high energy cosmic ray experiments. Namely, if $\tau_{\rm dec} \simeq {\rm few} \cdot 10^{26}$ sec, decaying DM can be invoked to explain the excesses in the fluxes of positrons and electrons measured by \PAMELA, \FERMI\ and \HESS, see Ref.~\cite{Cirelli:2012ut} and references therein. 
On the other hand, this value of $\tau_{\rm dec}$ is so much longer than the age of the Universe that the slow decay does not make a dent in the overall cosmological DM abundance and does not spoil the agreement with a number of astrophysical and cosmological observations~\cite{astroboundsdecayingDM}.

From the phenomenological point of view, the main feature of decaying DM with respect to the `more traditional' annihilating DM is that it is less constrained by neutral messenger probes (essentially $\gamma$-rays, but also neutrinos) originating from dense DM concentrations such as the galactic center, the galactic halo or nearby galaxies. The reason is simple and well-known: while the signal originating from annihilating DM is proportional to the square of the DM density, for decaying DM the dependence is on the first power; as a consequence, dense DM concentrations shine above the astrophysical backgrounds if annihilation is at play, but remain comparatively dim if DM is decaying. Decaying DM `wins' instead, generally speaking, when large volumes are considered. This is why in the following we will focus on targets as large as galaxy clusters or, essentially, the whole Universe.

On the observational side, the \FERMI\ and \HESS\  telescopes are making unprecedented progress in the field of  $\gamma$-ray astronomy, producing measurements of many different targets including those of interest for decaying DM. To this aim, we will compute the predicted signal from decaying DM, for a variety of decay channels and compare it to the $\gamma$-ray measurements, deriving constraints on the decay half-life.
Here, we make use of two distinct probes: (i) The isotropic residual $\gamma$-ray flux recently measured by \FERMI~\cite{FERMIexg}, 
which now extends from about 200 MeV up to 580 GeV. 
(ii)The recent observation in $\gamma$-rays of the Fornax galaxy cluster by \HESS~\cite{HESSFornax} and the sensitivity of the  upcoming large \v Cerenkov Telescope Array 
(\CTA)~\cite{CTA}. 
Our analysis presented here benefits from the release of new data on charged cosmic rays, the inclusion of EW corrections in all our computation of DM-generated fluxes, and an improved propagation scheme for $e^\pm$ in the Galaxy. See Ref.~\cite{Cirelli:2012ut} for more details.

The rest of this paper is organized as follows. In Sec.~\ref{charged fits} we update charged CR fits. 
In Sec.~\ref{isotropic} we discuss the calculation of the constraints from the isotropic residual $\gamma$-ray flux and those from the Fornax galaxy cluster.
In Sec.~\ref{results} we present the combined results.
In Sec.~\ref{conclusions} we present our conclusions.

\section{Update of the decaying DM fits to charged CR anomalies}
\label{charged fits}
The anomalous \PAMELA, \FERMI\ and \HESS\ data in $e^+$ and $(e^++e^-)$ have been interpreted in terms of DM decay. 
We use the following data sets:
(i) \PAMELA\ positron fraction~\cite{PAMELApositrons}; 
(ii) \FERMI\ positron fraction~\cite{FERMIpositrons};
(iii) \FERMI\ $(e^++e^-)$ total flux~\cite{Ackermann:2010ij}, provided in the low energy (LE) and high energy (HE) samples; 
(iv) \HESS\ $(e^++e^-)$ total flux~\cite{HESSleptons, HESSleptons2}, also provided in a lower energy portion and a higher energy one; 
(v) \MAGIC\ $(e^++e^-)$ total flux~\cite{Tridon:2011dk}; 
(vi) \PAMELA\ $\bar p$ flux~\cite{Adriani:2010rc}. 
We perform the fit to these data using the DM generated fluxes as provided in~\cite{PPPC4DMID}.
In looking for the best fitting regions, we scan over the propagation parameters of charged cosmic rays and over the uncertainties on the slope and normalization of the astrophysical electron, positron and antiproton background. We have assumed a NFW profile for the galactic DM halo. 
 \begin{figure*}[!t]
  \centering
 \mbox{ \includegraphics[width=0.45\textwidth]{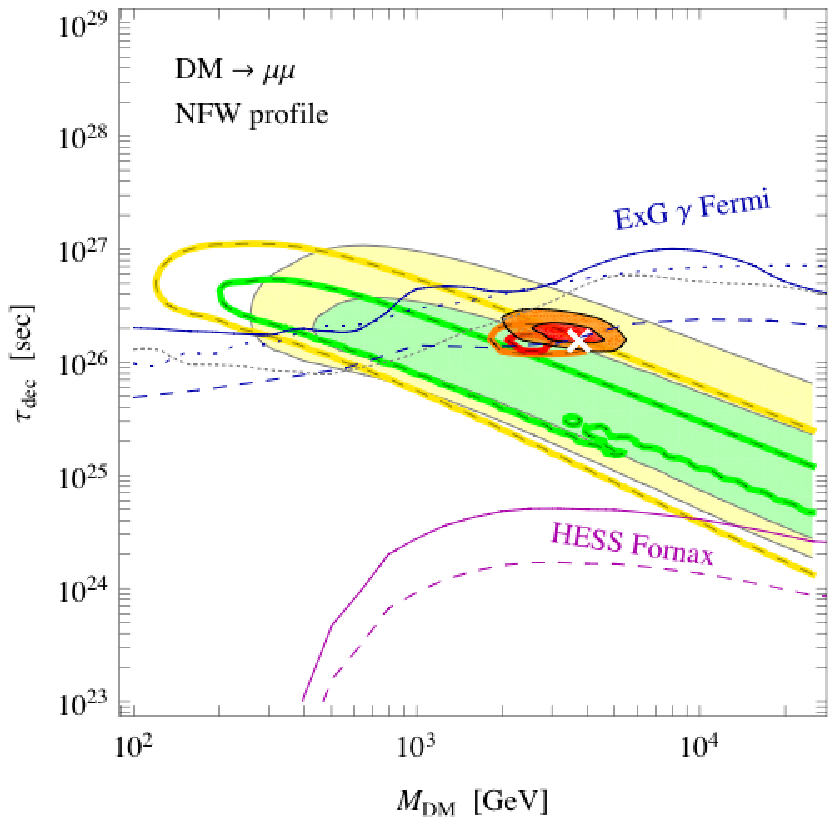}  \includegraphics[width=0.45\textwidth]{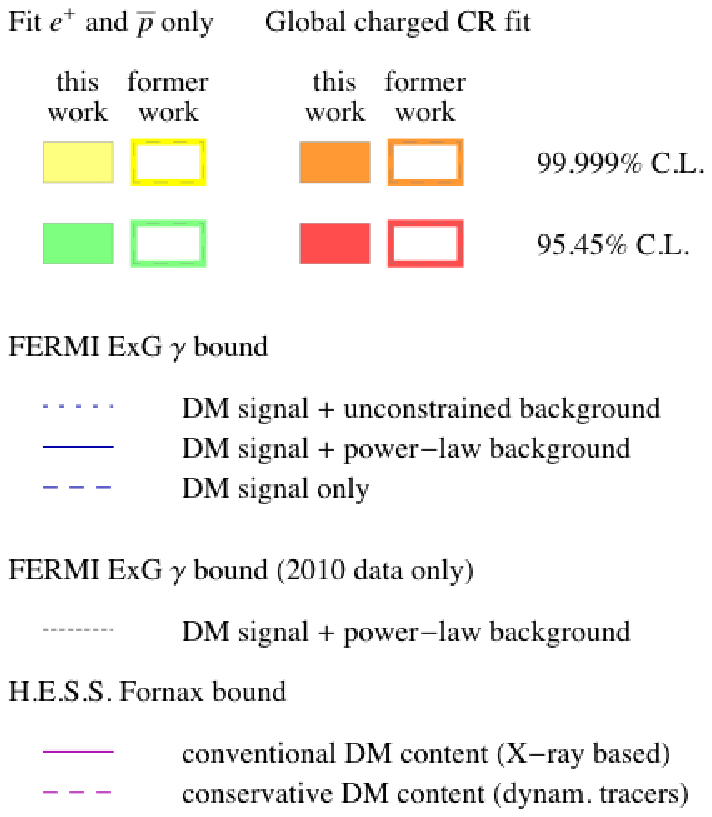}}
  \caption{Illustrative example (for the channel DM $\to \mu^+\mu^-$) of the impact on the fit and constrained regions 
following from different assumptions and choices, as discussed in the text.}
  \label{fig1}
 \end{figure*}

Positron fraction, e$^+$e$^-$ spectrum, antiprotron spectrum and the isotropic $\gamma$-ray flux from the above-mentioned datasets are fitted for various decaying channels. 
The typical decay time scales that are required for the global fit are of the order of $10^{26}$ to $10^{27}$ seconds. 
Only `leptophilic' channels allow a global fit: for the quark and gauge boson channels, the few TeV decaying DM needed by ($e^++e^-$) is in conflict with $\bar p$ data. See Ref.~\cite{Cirelli:2012ut} for more details.
The impact of the new data and the improved analysis tools on the identification of the best fit DM properties 
is shown in Fig.~\ref{fig1} for the $\mu^+\mu^-$ channel.


\section{Isotropic $\gamma$-ray flux}
\label{isotropic}
The measurements in~\cite{FERMIexg} by the \FERMI\ satellite correspond to the (maximal) residual, isotropic $\gamma$-ray flux present in their data. Its origin can be in a variety of different phenomena, both in the form of unresolved sources and in the form of truly diffuse processes (see~\cite{FERMIexg} and reference therein). 
DM decays can also contribute to this isotropic flux, with two terms: 1) an extragalactic cosmological flux, due to the decays at all past redshifts; 2) the residual emission from the DM halo of our Galaxy.
The former is of course truly isotropic, at least as long as one neglects possible nearby DM overdensities. The latter is not, but its minimum constitutes an irreducible contribution to the isotropic flux. 
In formul\ae, the predicted differential DM flux that we compare with \FERMI\ isotropic diffuse $\gamma$-ray data is therefore given by 
\begin{equation}  
\frac{d \Phi_{\rm isotropic}}{d E_\gamma} = \frac{d \Phi_{\rm ExGal}}{d E_\gamma} + 
4 \pi \left. \frac{d \Phi_{\rm Gal}}{d E_\gamma\, d \Omega}\right|_{\rm minimum}
\label{eq:main}
\end{equation}  

The extragalactic flux is given, in terms of the Earth-measured photon energy $E_\gamma$, by 
\begin{equation}
\frac{d \Phi_{\rm ExGal}}{d E_\gamma} = \Gamma_{\rm dec}\, \frac{\Omega_{\rm DM}\,\rho_{c,0}}{M_{\rm DM}} 
\,\int_0^\infty d z\,\frac{e^{-\tau(E_\gamma(z),z)}}{H(z)} \frac{d N}{d E_\gamma}(E_\gamma(z),z)\,,
\label{smoothedmap}
\end{equation}
where $\Gamma_{\rm dec}= \tau_{\rm dec}^{-1}$ is the decay rate.
Here the Hubble function $H(z) = H_0\sqrt{\Omega_M(1+z)^3+\Omega_\Lambda}$, where $H_0$ is the present Hubble expansion rate. $\Omega_{\rm DM}$, $\Omega_M$ and  $\Omega_\Lambda$ are respectively the DM, matter and cosmological constant energy density in units of the critical density, $\rho_{c,0}$. 
The $\gamma$-ray spectrum $d N/d E_\gamma$, at any redshift $z$, is the sum of two components: (i) the high energy contribution due to the prompt $\gamma$-ray emission from DM decays and (ii) the lower energy contribution due to Inverse Compton Scatterings (ICS) on CMB photons of the $e^+$ and $e^-$ from those same decays. 
Using Eq.~(\ref{smoothedmap}), the extragalactic flux can therefore be computed in terms of known quantities for any specified DM mass $M_{\rm DM}$ and decay channel. 
The factor $e^{-\tau(E_\gamma,z)}$ in Eq.~(\ref{smoothedmap}) accounts for the absorption of high energy $\gamma$-rays due to scattering with the extragalactic UV background light. 

The flux from the galactic halo, coming from a generic direction $d \Omega$, is given by the well known expression 
\begin{eqnarray}
\frac{d \Phi_{\rm Gal}}{d E_\gamma \ d \Omega} = \frac{1}{4\pi} \frac{\Gamma_{\rm dec}}{M_{\rm DM}} \int_{\rm{los}} d s \, \rho_{\rm halo}[r(s,\psi)] \, \frac{d N}{d E_\gamma}\,,
\label{fluxdec}
\end{eqnarray}
i.e. as the integral of the decaying DM density piling up along the line of sight (los) individuated by the direction $d \Omega$. $\rho_{\rm halo}$ is the DM distribution in the Milky Way, for which we take a standard NFW~\cite{Navarro:1995iw} profile 
\begin{eqnarray}
 \rho_{\rm NFW}(r)=\rho_{s}\frac{r_{s}}{r}\left(1+\frac{r}{r_{s}}\right)^{-2},
   \label{eq:NFW}
\end{eqnarray}
with parameters $r_{s} = 24.42$ kpc and $\rho_{s} = 0.184$ GeV/cm$^3$~\cite{PPPC4DMID}. The coordinate $r$, centered on the GC, reads $r(s,\psi)=(r_\odot^2+s^2-2\,r_\odot\,s\cos\psi)^{1/2}$, where $r_\odot = 8.33$ kpc and $\psi$ is the angle between the direction of observation in the sky and the GC.
As indicated in Eq.~(\ref{eq:main}), we need to determine the minimum of the flux in Eq.~(\ref{fluxdec}). We choose to locate the minimum always at the anti-GC. 

$d \Phi_{\rm isotropic}/d E_\gamma$ is compare with the \FERMI\ data of~\cite{FERMIexg}. 
The DM signal does not agree in shape with the data, which are instead well fit by a simple power law~\cite{FERMIexg}, and we are driven to derive constraints on the maximum DM signal, and therefore the minimum $\tau_{\rm dec}$, admitted by the data. 

\begin{figure*}[!t]
  \centering
 \mbox{ \includegraphics[width=0.4\textwidth]{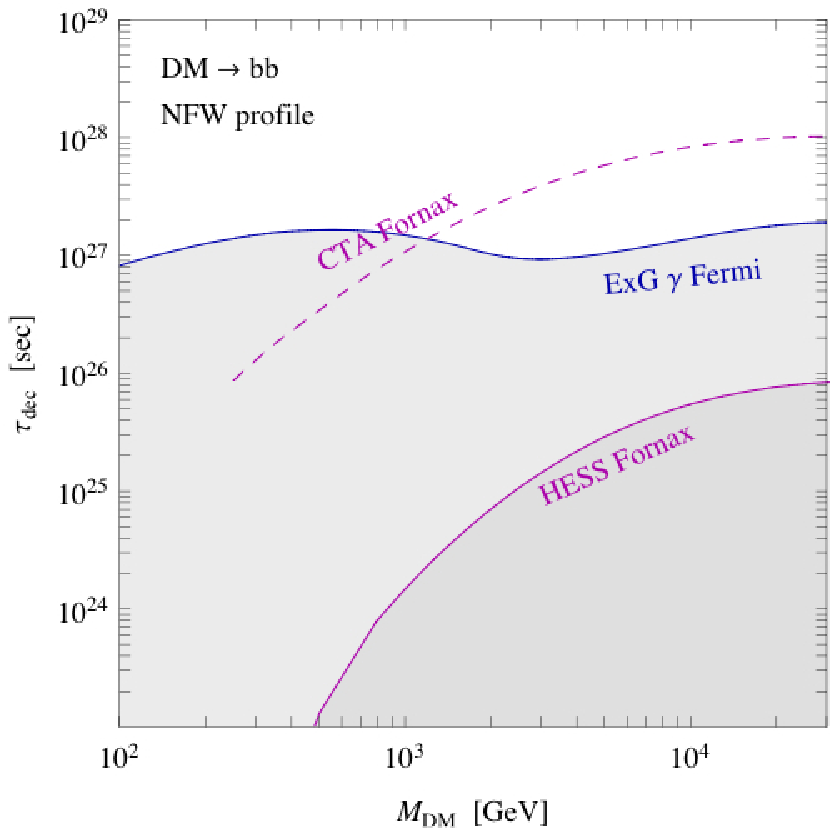}  \includegraphics[width=0.4\textwidth]{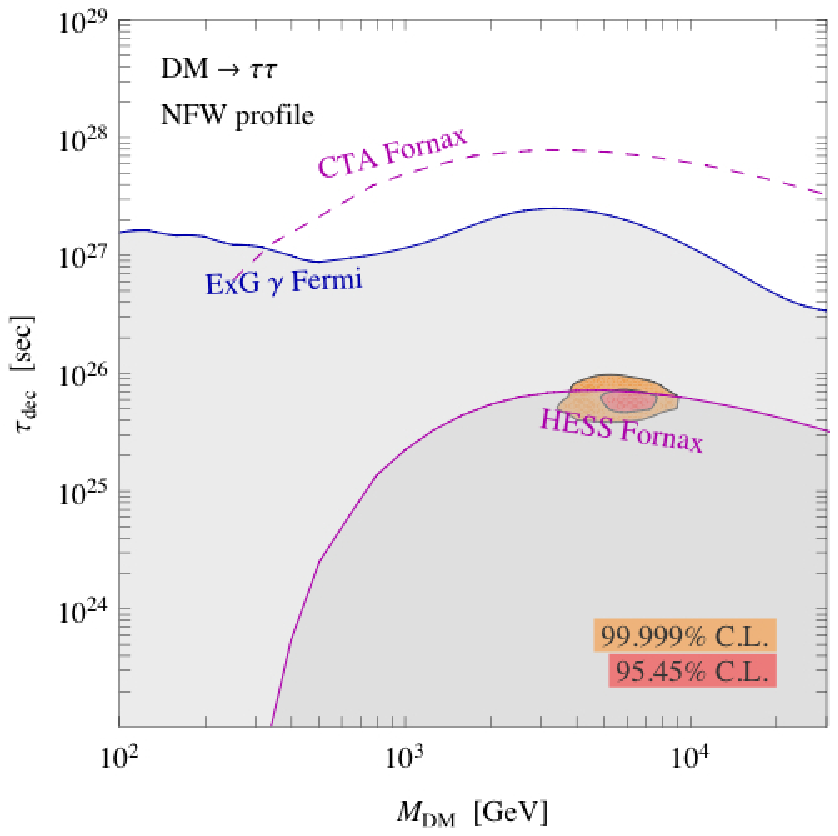}}
  \caption{The regions on the parameter space M$_{\rm DM}$ - $\tau_{\rm dec}$ that are excluded by the \FERMI\ and \HESS\ constraints and 
that can be explored by \CTA, together with the regions of the global fit to the charged CR data, for $b\bar{b}$ and $\tau^+\tau^-$ decay channels.}
  \label{fig2}
 \end{figure*}

There are however several possible ways to compute such constraints. Our procedure uses {\em `DM signal + power-law background'} to obtain the fiducial constraints shown in 
Fig.~\ref{fig1}. This procedure is `fiducial' also in the sense that it matches the analysis we do for charged CR anomalies (see Sec.~\ref{charged fits}) and therefore fit regions and constraints are essentially consistent with each other in Fig.~\ref{fig2}. For discussion on this procedure and alternative ones, see ~\cite{Cirelli:2012ut}. 

\section{Fornax cluster $\gamma$-ray flux}
\label{Fornax}

Galaxy clusters are the largest gravitationally bound structures in the universe,  80\% of their total mass being in the form of DM. 
Although they are located at much larger distances than other popular targets such as dwarf satellites of the Milky Way, they turn out to be attractive environments to search for DM due to their predicted high DM luminosity. A few galaxy clusters have been observed by \HESS: the most attractive of them for DM searches is Fornax~\cite{HESSFornax}.

The predicted DM $\gamma$-ray flux from Fornax can be easily obtained by integrating Eq.(\ref{fluxdec}) over the observational solid angle $\Delta\Omega$, and replacing $\rho_{\rm halo}$ with $\rho_{\rm Fornax}$. Various DM halo models for Fornax have been considered in~\cite{HESSFornax} and 
varying over these profiles, the difference in the flux factor is less than a factor of 3 for a given opening integration angle. We take as `fiducial' the X-ray based 
determination of the DM profile, hereafter referred to as the RB02 profile~\cite{HESSFornax}. 

Decaying DM searches largely benefit from an optimization of the opening angle to guarantee the highest signal-to-noise ratio.  As it is straightforward from Eq.(\ref{fluxdec}), the luminosity scales  with the size of the solid integration angle. On the other hand, background is increasing as well. We find that the optimization of the signal-to-noise ratio 
for the RB02 profile yields a solid angle of $\rm \Delta\Omega = 2.4\times10^{-4}$ sr.          

In order to extract exclusion limits on the DM lifetime from $\gamma$-ray astronomical observation with IACTs, the background needs to be determined to constrain the decaying DM luminosity. The background is calculated in a region referred to as the OFF region, and the signal region as the ON region.
Both ON and OFF regions depend on the observation mode and are specific to the IACT instrument.  As for the background level, it is taken simultaneously, {\it i.e.} in the same data-taking observing conditions, to the signal events in order to allow for the most accurate estimate in the so-called {\it template} method  procedure~\cite{HESSFornax}.
\HESS\ has observed Fornax for a total of 14.5 hours at low zenith angle to allow for best sensitivity to low DM masses.

\section{Results and discussion}
\label{results}
Figure~\ref{fig2} presents the exclusion plots for $b\bar{b}$ and $\tau^+\tau^-$ channels. One can see that the constraints from the \FERMI\ isotropic $\gamma$-ray data exclude decaying DM with a lifetime of a few $\times 10^{27}$ seconds. They rule out the charged CR fit regions for the $b\bar{b}$ channel. As illustrated in the example in Fig.~\ref{fig1}, adopting the more conservative constraint procedure may marginally reallow a portion of the fit regions, for the DM $\to \mu^+\mu^-$, but leaving a clear tension.
Keeping only data published in 2010 in~\cite{Abdo:2010nz} still allows to exclude the CR fit regions. The constraints from \FERMI\ rise gently as a function of the mass, essentially as a consequence of the fact that the measured flux rapidly decreases with energy. They also depend (mildly, a factor of a few at most) on the decay channel~\cite{Cirelli:2012ut}. 
The constraints from \HESS\ Fornax remain subdominant, roughly one order of magnitude below the \FERMI\ ones. However, for the case of the DM $\to \tau^+\tau^-$ channel, the bound also reaches the CR fit region and essentially confirms the exclusion (Fig.~\ref{fig2}). They do not look competitive with respect to \FERMI\ even for larger masses.

With respect to the isotropic $\gamma$-ray constraints of~\cite{CPS1}, the bounds derived here are stronger by a factor of 2 to 3. The reasons of this essentially amount to: updated datasets, more refined DM analyses and the adoption of a more realistic constraint procedure. With respect to the work in~\cite{Huang:2011xr}, our constraints from the \FERMI\ isotropic background are somewhat stronger than their corresponding ones. 
In addition,~\cite{Huang:2011xr} presents bounds from the observation of the Fornax cluster by \FERMI: these are less stringent than our \FERMI\ isotropic background constraints but more powerful than our \HESS-based constraints at moderate masses. At the largest masses, our \HESS-based constraints pick up and match theirs (for the $b \bar b$ channel).The bounds from clusters other than Fornax are less powerful, according to~\cite{Huang:2011xr}.  
On the other hand, the preliminary constraints shown (for the $b\bar b$ channel only) in~\cite{Zimmer:2011vy}, obtained with a combination of several clusters in \FERMI, exceed our bounds by a factor of 2. 
Bounds from probes other than the isotropic flux and clusters do not generally achieve the same constraining power. 

Within the context of observations performed by IACT, we note that the decay lifetime constraints obtained with galaxy clusters are stronger than those from dwarf galaxies. Even for the ultra-faint dwarf galaxy Segue 1, which is believed to be the most promising dwarf in the northern hemisphere,
the constraints are 2 orders of magnitude higher for a DM particle mass of 1 TeV~\cite{Aliu:2012ga}.
We also estimate that the constraints that we derive are stronger than those that can come from neutrino observation by \ICECUBE\ of the Galactic Center (see for instance~\cite{Abbasi:2011eq}). In summary: with the possible exception of the preliminary bounds from a combination of clusters by \FERMI\ for the $b \bar b$ channel, the constraints that we derive from the isotropic $\gamma$-ray flux are the most stringent to date.

Currently the constraints on decaying DM from the \FERMI\ satellite are dominant with respect to those from \HESS. However, while the former may increase its statistics by at most a factor of a few, for the latter there are prospects of developments in the mid-term future.
The next-generation IACT will be a large array composed of a few tens to a hundred telescopes~\cite{Consortium:2010bc}. The goal is to improve the overall performances of the present generation: one order-of-magnitude increase in sensitivity and enlarge the accessible energy range both towards the lower and higher energies allowing for an energy threshold down to a few tens of GeV.  From the actual design study of \CTA, the effective area will be increased by a factor $\sim$10 and a factor 2 better in the hadron rejection is expected. 
The calculation of   95\% C.L. \CTA\ sensitivity is detailed in~\cite{Cirelli:2012ut} and given by 
\begin{eqnarray*}
\Gamma_{\rm dec}^{\rm 95\% C.L.} = \frac{4\pi}{\int_{\Delta\Omega} d\Omega \int_{\rm los} ds \, \rho_{\rm Fornax}[r(s,\psi)]} \\
 \times \frac{M_{\rm DM} \,  N_{\gamma}^{\rm 95\% C.L.}}{T_{\rm obs}\, \int_{0}^{M_{\rm DM}/2} A_{\rm CTA}(E_{\gamma}) \, \frac{dN}{d E_{\gamma}}(E_\gamma) \, d E_{\gamma}} \, ,
\end{eqnarray*}
where $ N_{\gamma}^{\rm 95\% C.L.}$ is the limit on the number of $\gamma$-ray events, $A_{\rm CTA}$ is the CTA effective area and $T_{\rm obs}$ the observation time.
Figure~\ref{fig2} shows the 95\% C.L. sensitivity of \CTA\ on the decay lifetime for the RB02 halo profile for 50h observation time and $\rm \Delta\Omega = 2.4 \times 10^{-4}$ sr.
To conclude, we also mention that a technique which could allow for significant improvements in the exploration of the parameter space of decaying DM using clusters is the one of stacking the observation of a large number of different clusters, as recently discussed in~\cite{Combet:2012tt}. 
The authors find that improvements of up to 100 can be theoretically achieved, albeit this factor is $\sim$5 for more realistic background-limited instruments.


\section{Conclusions}
\label{conclusions}
Decaying DM has come to the front stage recently as an explanation, alternative to annihilating DM, for the anomalies in CR cosmic rays in \PAMELA, \FERMI\ and \HESS. But, more generally, decaying DM is a viable possibility that is or can naturally be embedded in many DM models. It is therefore interesting to explore its parameter space in the light of the recent observational results.
We discussed the constraints which originate from the measurement of the isotropic $\gamma$-ray background by \FERMI\ and of the Fornax cluster by \HESS, for a number of decaying channels and over a range of DM masses from 100 GeV to 30 TeV. We improved the analysis over previous work by using more recent data and updated computational tools.
We found that the constraints by \FERMI\ rule out decaying half-lives of the order of $10^{26}$ to few $10^{27}$ seconds. These therefore exclude the decaying DM interpretation of the charged CR anomalies, for all 2-body channels, at least adopting our fiducial constraint procedure. The constraints by \HESS\ are generally subdominant. For the DM $\to \tau^+ \tau^-$ channel, they can however also probe the CR fit regions and essentially confirm the exclusion.
With one possible exception for the DM $\to b \bar b$ channel, the constraints that we derive from the isotropic $\gamma$-ray flux are the most stringent to date.
We also discussed the prospects for the future \v Cerenkov telescope \CTA, which will be able to probe an even larger portion of the parameter space.


\end{document}